\documentclass[
twocolumn,
superscriptaddress,
showpacs,
nofootinbib,
nobibnotes,
amsmath,amssymb,
aps,
 pre,
floatfix,
]{revtex4}

\usepackage{graphicx}% Include figure files
\usepackage{dcolumn}% Align table columns on decimal point
\usepackage{bm}% bold math
\usepackage{natbib}
\usepackage[caption=false]{subfig}
\def \be {\begin{equation}}
\def \ee {\end{equation}}

\def\dbar{{\mathchar'26\mkern-12mu d}} 
\begin{document}

\preprint{APS/123-QED}

\title{Quantum correlated heat engine with nonlinear spin-spin interactions}
%
%-------------------------------------------------------------------------------
\author{Ferdi Altintas}
\affiliation{Department of Physics, Abant Izzet Baysal University, Bolu, 14280, Turkey}
\author{Ali \"U. C. Hardal}
%\affiliation{Department of Physics, Ko\c{c} University, \.Istanbul, Sar{\i}yer,, 34450, Turkey}
%
\author{\"{O}zg\"{u}r E. M\"{u}stecapl{\i}o\u{g}lu}
\email{omustecap@ku.edu.tr}
\affiliation{Department of Physics, Ko\c{c} University, \.Istanbul, Sar{\i}yer, 34450, Turkey}
%
%-------------------------------------------------------------------------------
\date{\today}
%-------------------------------------------------------------------------------
\begin{abstract}
We propose a four level quantum heat engine in Otto cycle with a working substance of 
two spins subject to an external magnetic field and coupled to each other by a one-axis twisting spin squeezing 
nonlinear interaction. We calculate the positive work and the efficiency of the engine for different parameter regimes. In particular, we investigate the effects of quantum correlations at the end of the two isochoric processes of the Otto cycle, as measured by the entanglement of formation and quantum discord, on the work extraction and efficiency. The regimes where the quantum correlations could enhance the efficiency and work extraction are characterized.
\end{abstract}
\pacs{05.30.-d, 05.70.-a, 03.65.Ud, 03.67.Mn}
%
%03.67.Mn	Entanglement measures, witnesses, and other characterizations
%03.65.Ud	Entanglement and quantum nonlocality
%05.70.-a	Thermodynamics
%05.30.-d	Quantum statistical mechanics 
\maketitle
%%%%%%%%%%%%%%%%%%%%%%%%%%%%%%%%%%%%%%%%%%%%%%%%%%%%%%%%%%%%%%%%%%%%%%%%%%%%%%%%%%%%%%%%%%%%%%%%%%%%%%%%%%%%%%%%%%%%%
%%%%%%%%%%%%%%%%%%%%%%%%%%%%%%%%%%%%%%%%%%%%%%%%%%%%%%%%%%%%%%%%%%%%%%%%%%%%%%%%%%%%%%%%%%%%%%%%%%%%%%%%%%%%%%%%%%%%%
\section{Introduction}
Heat engines are crucial tools for our modern society and their miniaturization is required for our further development beyond the industrial era.
Recent progress in producing and controlling systems in micro, nano and even in atomic length scales 
inspired many proposals of heat engines which could operate in quantum realm. 
Such engines are called as quantum heat engines (QHEs) and despite
their small length scales they promise surprisingly high efficiency ~\cite{RevModPhys.81.1,PhysRevLett.104.207701,PhysRevE.72.066118,PhysRevA.88.013842,PhysRevE.71.046106,PhysRevE.85.041148,
kieu04,quan07,huang12,wang09}. Some intriguing QHEs are quantum information engines which could exploit quantum coherence as a resource to harvest useful work more efficiently than the Carnot limit, without violating the second law of thermodynamics~\cite{PhysRevLett.104.207701,huang12}. In addition to the appealing practical value of implementing such miniature quantum machines, their fundamental studies could extend thermodynamics to quantum regime and establish its place there from the perspective of quantum information theory.

We propose here a QHE with a working substance of a pair of spins subject to an external magnetic field and coupled to each other  by a so called one-axis twisting nonlinear spin squeezing interaction~\cite{kitagawa1993squeezed,wang03,Ma201189}. The motivation behind considering this particular interaction is its capability to establish pairwise quantum correlations among the spins. The engine is assumed to operate in quantum Otto cycle~\cite{kieu04,quan07} that consists of two adiabatic and two isochoric stages. Our first objective is to examine the work extraction out of the engine and its efficiency 
by changing either the external field or spin-spin interaction strength in the adiabatic stages. 

Our second objective is to investigate how the quantum correlations play a role in the performance of the engine. The quantum state of the working medium at the end of the two isochoric processes can be represented by a canonical ensemble. Since the temperature of the heat baths as well as the system parameters can be externally controlled~\cite{gross2010nonlinear,riedel2010atom}, it would also be possible to control the amount of quantum correlations of the working substance in the two heat baths. We discuss the quantum correlations in our system by analysing the behavior of entanglement of formation (EoF)~\cite{PhysRevLett.80.2245} and quantum discord (QD)~\cite{ollivier01,henderson01}. QD can measure quantum correlations beyond entanglement (or "quantumness") and more robust than entanglement in the presence of thermal noise~\cite{modi13}. We would like to explore if quantum correlations and entanglement are mere byproducts of the QHE or if they can enhance its performance. The general answer to this question remained to be elusive so far and model dependent effects are reported, as we shall summarize below. Our choice of spin squeezing nonlinear interaction model is on purpose to establish strong bipartite quantum correlations and entanglement in the QHE. 

Along similar lines to ours, quantum entanglement was studied as a quantum resource in QHEs~\cite{huang12,zhang08,zhang07,wang09,hovhan13,brunner13,alici13}; and its positive or negative effects on the engine performance is found to be model dependent. In particular, work cannot be extracted in some coupled spin models, if the entanglement of spins in the hot bath exceeds the one in the cold bath~\cite{huang12,zhang08,zhang07,wang09}. Conversely, entanglement could enhance the performance of some quantum machines~\cite{brunner13,alici13,park13,funo13}. QD was investigated in the context of quantum absorption chillers for a particular model of interacting three qubits (two level systems); albeit its role on the performance of the quantum refrigerator remained elusive and unclear~\cite{correa13}. 

Our analysis revealed the parameter regimes in our model for which QD or entanglement are not mere byproducts of interactions but they can be used as a resource. We find that entanglement of the working substance at the end of hot bath stage of the Otto cycle forbids the work extraction, if the external field or the interactions are stronger in hot bath stage relative to cold bath stage. In this case, QD at the end of cold bath stage is constructive for positive work and efficiency over a wide range of interaction strengths; while QD at the hot bath stage can be either constructive or destructive depending on the system parameters. Their difference is constructive provided that QD  is greater at the cold bath stage. When the interaction is weaker in the hot bath stage, both the entanglement and QD lead to enhanced work and efficiency. 

This article is organized as follows. In Sec.~\ref{sec:model}, we describe our two spin QHE by describing the working substance, its thermalization and the Otto cycle. In Sec.~\ref{sec:theory} we summarize the quantum work, efficiency and quantum correlation measures which we calculate. The results are reported and discussed in Sec.~\ref{sec:results}. We present our conclusions in Sec.~\ref{sec:conc}.
%%%%%%%%%%%%%%%%%%%%%%%%%%%%%%%%%%%%%%%%%%%%%%%%%%%%%%%%%%%%%%%%%%%%%%%%%%%%%%%%%%%%%%%%%%%%%%%%%%%%%%%%%%%%%%%%%%%%%%
%%%%%%%%%%%%%%%%%%%%%%%%%%%%%%%%%%%%%%%%%%%%%%%%%%%%%%%%%%%%%%%%%%%%%%%%%%%%%%%%%%%%%%%%%%%%%%%%%%%%%%%%%%%%%%%%%%%%%%
\section{Model System: The Quantum Otto Engine}\label{sec:model}
%%%%%%%%%%%%%%%%%%%%%%%%%%%%%%%%%%%%%%%%%%%%%%%%%%%%%%%%%%%%%%%%%%%%%%%%%%%%%%%%%%%%%%%%%%%%%%%%%%%%%%%%%%%%%%%%%%%%%%
\subsection{Working Substance}\label{sec:wsubs}
%%%%%%%%%%%%%%%%%%%%%%%%%%%%%%%%%%%%%%%%%%%%%%%%%%%%%%%%%%%%%%%%%%%%%%%%%%%%%%%%%%%%%%%%%%%%%%%%%%%%%%%%%%%%%%%%%%%%%%
We consider a four level quantum Otto engine with a working medium of two spins under an external magnetic field and coupled to each other by a one-axis twisting spin squeezing interaction. The Hamiltonian of the working substance can be written as~\cite{kitagawa1993squeezed,wang03}
\begin{eqnarray}\label{hamilton}
H=\mu S_x^2+\Omega S_z,
\end{eqnarray}
where $\Omega$ ($\geq0$) is the strength of the external magnetic field in $z$ direction, $\mu$ ($\geq0$)  describes the strength of the spin squeezing interaction in $x$ direction, $S_{\alpha}=\frac{1}{2}\sum_{i=1}^2\sigma_{\alpha}^i, (\alpha\in\{x,y,z\})$ are the collective spin operators, and $\sigma_{\alpha}^i$ are the Pauli matrices for the $i^{th}$ spin. The interaction establishes pairwise correlations between all the individual spins in the collective spin system; in our case bipartite correlations are formed between the two spins.

\begin{figure}[ht!]
\includegraphics[width=8.0cm]{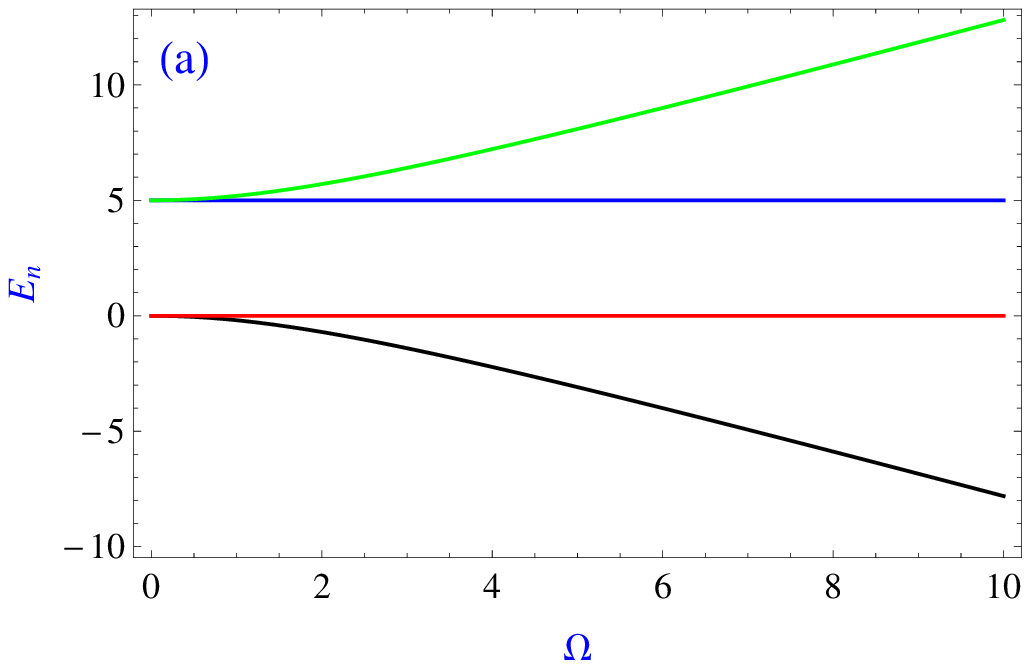}
\includegraphics[width=8.0cm]{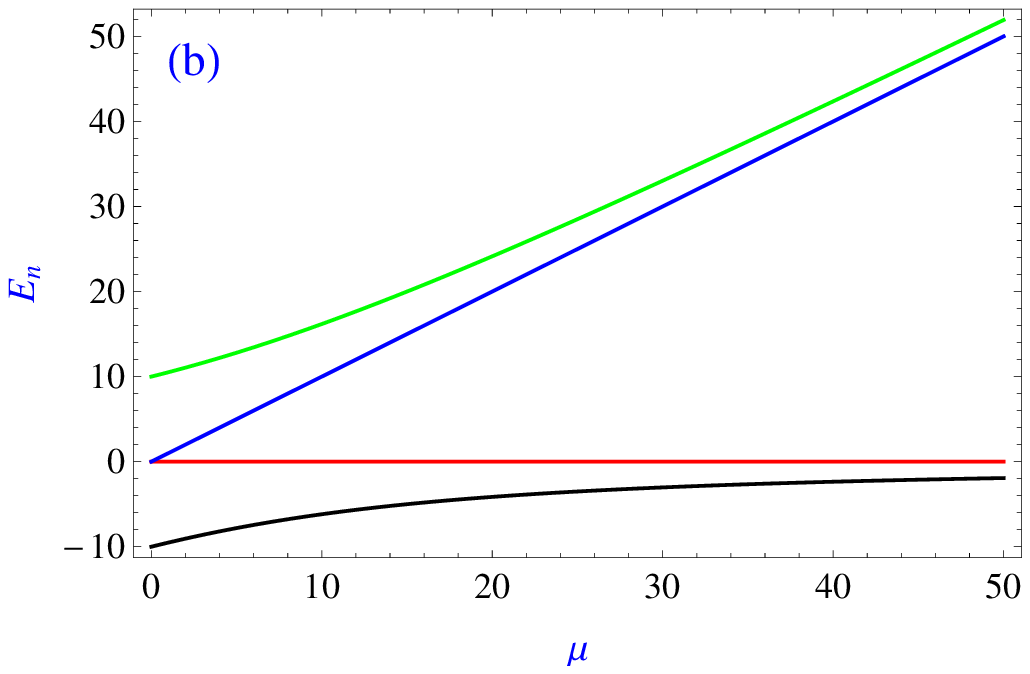}
\caption{\label{fig1} The energy spectrum $E_n$ ($n=1,2,3,4$ from bottom to top, respectively) versus $\Omega$ for $\mu=5$ (a), and versus $\mu$ for $\Omega=10$ (b).}
\end{figure}

For a system of two spins, the eigenvalues $E_{n}$ and the corresponding eigenvectors $|\Psi_{n}\rangle$ of the Hamiltonian~(\ref{hamilton}) can be calculated as:
\begin{eqnarray}\label{spectrum}
E_{1,4}&=&\frac{\mu\mp \kappa}{2}, \quad \left|\Psi_1\right\rangle=\frac{1}{A_\mp}\left(\mu\left|00\right\rangle+(2\Omega\mp\kappa)\left|11\right\rangle\right),\nonumber\\
E_2&=&0,\quad\quad\quad \left|\Psi_2\right\rangle=\frac{1}{\sqrt{2}}\left(\left|10\right\rangle-\left|01\right\rangle\right),\nonumber\\
E_3&=&\mu,\quad\quad\quad  \left|\Psi_3\right\rangle=\frac{1}{\sqrt{2}}\left(\left|10\right\rangle+\left|01\right\rangle\right),
%\nonumber
%E_4&=&\frac{\mu+\kappa}{2}, \quad \left|\Psi_4\right\rangle=\frac{1}{A_+}\left(\mu\left|00\right\rangle+(2\Omega+\kappa)\left|11\right\rangle\right),\nonumber\\
\end{eqnarray}
where $\kappa=\sqrt{\mu^2+4\Omega^2}$ and $A_{\pm}=\sqrt{\mu^2+(2\Omega\pm\kappa)^2}$. The eigenvalues are plotted in Fig.~\ref{fig1}. 
%
%%%%%%%%%%%%%%%%%%%%%%%%%%%%%%%%%%%%%%%%%%%%%%%%%%%%%%%%%%%%%%%%%%%%%%%%%%%%%%%%%%%%%
\subsection{Thermalization}\label{sec:thermal}
%%%%%%%%%%%%%%%%%%%%%%%%%%%%%%%%%%%%%%%%%%%%%%%%%%%%%%%%%%%%%%%%%%%%%%%%%%%%%%%%%%%%%%%%%%%%%%
When thermal fluctuations are introduced into the system, the reduced density matrix of the working substance at thermal equilibrium with the heat bath can be written as:
\begin{eqnarray}\label{thermalden}
\rho&=&\frac{1}{Z}e^{-\beta H}\nonumber\\
&=&\sum_{n=1}^{4}P_{n}(T)\left|\Psi_n\right\rangle\left\langle \Psi_n\right|,
\end{eqnarray}
where $P_{n}(T)=e^{-\beta E_n}/Z$ are the occupation probabilities of the eigenstates, $\beta=1/k_{B}T$ ($k_{B}=1$), $T$ is the temperature, and $Z=\sum_n e^{-\beta E_n}$ is the partition function.
%%%%%%%%%%%%%%%%%%%%
\subsection{Quantum Otto cycle}\label{sec:cycle}
%%%%%%%%%%%%%%%%%%%%
The stages of the proposed quantum Otto engine can be described as follows:\\
\textit{Step 1}: The working substance, having energy levels $E_n^H$ and initial probabilities in each eigenstates, is brought into contact with a hot heat bath at $T =T_H$.  Upon attaining  thermal equilibrium with the heat bath, the occupation probabilities change to $P_n(T_H)$, while the energy levels remains the same. An amount of heat is absorbed, but no work is done during this quantum isochoric process.\\
\textit{Step 2}: The working medium is isolated from the heat bath and exposed to quantum adiabatic expansion process, in which the energy structure is changed from $E_n^H$ to $E_n^L$. Provided the change of the energies is slow enough, the occupation probabilities are maintained according to the quantum adiabatic theorem. An amount of work is done, but no heat is exchanged.\\
\textit{Step 3}: The working substance goes through another quantum isochoric process where it is brought into contact with a cold heat bath at $T=T_L$ ($T_H > T_L$).  After the thermalization process, an amount of heat is released but no work is done.  At the end of the stage, the occupation probabilities become $P_n (T_L)$, while the energy levels, $E_n^L$, remains unchanged.\\
\textit{Step 4}: The system is removed from the cold heat bath and undergoes quantum adiabatic contraction process which changes the energy levels from  $E_n^L$ to $E_n^H$, but keeps the probabilities the same.  An amount of work is done during the stage, but no heat is exchanged. At the end of this stage, the working medium returns to its initial condition and is ready to cycle again.
%%%%%%%%%%%%%%%%%%%%%%%%%%%%%%%%
\section{Theory} \label{sec:theory}
%%%%%%%%%%%%%%%%%%%%%%%%%%%%%%%%
\subsection{Work and efficiency}
The heat transfer and the work performed at the quantum level can be calculated through the interpretation of the quantum first law of thermodynamics~\cite{kieu04,quan07}: $dU=\dbar Q+\dbar W=\sum_n\{E_ndP_n+P_ndE_n\}$. In this interpretation, the infinitesimal heat transfer, $\dbar Q=\sum_nE_ndP_n$, is associated with the change of occupation probabilities and the infinitesimal work done, $\dbar W=\sum_nP_ndE_n$, is with the change in energy levels. The heat absorbed, $Q_{in}$, during the stage 1, heat released, $Q_{out}$, during the stage 3, the net work done, $W$, and the operational efficiency, $\eta$ of the engine can be obtained easily through this interpretation~\cite{kieu04,quan07}
\begin{eqnarray}\label{hwe}
Q_{in}&=&\sum_nE_n^H\left[P_n(T_H)-P_n(T_L)\right],\nonumber\\
Q_{out}&=&\sum_nE_n^L\left[P_n(T_L)-P_n(T_H)\right],\nonumber\\
W&=&Q_{in}+Q_{out}\nonumber\\
&=&\sum_n\left[E_n^H-E_n^L\right]\left[P_n(T_H)-P_n(T_L)\right],\nonumber\\
\eta&=&\frac{W}{Q_{in}},
\end{eqnarray}
where $E_n^H$ ($E_n^L$) are the energy levels during the stage 1 (3) which can be obtained by replacing $\mu$ and $\Omega$ by $\mu_H$ ($\mu_L$) and $\Omega_H$ ($\Omega_L$), respectively. For the positive work extraction ($W>0$), the physically acceptable situation is considered, i.e., $Q_{in} >-Q_{out}>0$. The possible cases,  $Q_{in}>Q_{out}>0$ and  $Q_{out}>-Q_{in}>0$, which violate the second law of thermodynamics are excluded in the present study. Moreover, we have considered the efficiency of the engine when $W>0$.

We will discuss the work extraction and efficiency by considering two different schemes for the the adiabatic steps: (a) the magnetic field is altered between two values $\left(\Omega_H\rightarrow\Omega_L\rightarrow\Omega_H\right)$ at a fixed interaction coefficient $\mu$, and (b) the squeezing coefficient is changed between two values $\left(\mu_H\rightarrow\mu_L\rightarrow\mu_H\right)$ at a fixed magnetic field $\Omega$. In both cases, we will also discuss the possibility of work extraction in two parameter regions (i) $\Omega_H>\Omega_L$ and (ii) $\Omega_H<\Omega_L$ for the case (a), and (i) $\mu_H>\mu_L$ and (ii) $\mu_H<\mu_L$ for the case (b). 
\subsection{Quantum discord and entanglement}
We will also study the role of quantum correlations as measured by quantum discord and entanglement on the thermodynamic quantities for these cases. 
First, we note that the thermal density matrix in the standard basis $\{\left|1\right\rangle\equiv\left|11\right\rangle,\left|2\right\rangle\equiv\left|10\right\rangle, \left|3\right\rangle\equiv\left|01\right\rangle, \left|4\right\rangle\equiv\left|00\right\rangle\}$ given in Eq.~(\ref{thermalden}) has matrix elements in the following form:
\begin{eqnarray}\label{denmatst}
\rho=\left(\begin{array}{cccc}
a & 0 & 0 & w\\
0 & b & z & 0\\
0 & z & b & 0\\
w & 0 & 0 & d\
\end{array}\right).
\end{eqnarray}
The matrix elements can be easily determined by using Eqs.~(\ref{spectrum}) and~(\ref{thermalden}). The calculation of QD requires a maximization procedure over projective measurements performed locally on one system, which cannot be done analytically in general. However, for the above density matrix, the extremization in the definition of quantum discord can be done analytically~\cite{fanchini10}. The analytic form of the QD can be written as~\cite{fanchini10}:
\begin{eqnarray}\label{qd}
D=\min\left\{D_1,D_2\right\}
\end{eqnarray}
where
\begin{eqnarray}\label{d1d2}
D_1&=&S(\rho_A)-S(\rho) - a\log_2\left(\frac{a}{a+b}\right)-b\log_2\left(\frac{b}{a+b}\right)\nonumber\\
   &-&d\log_2\left(\frac{d}{b+d}\right)-b\log_2\left(\frac{b}{d+b}\right),\nonumber\\
D_2&=&S(\rho_A)-S(\rho)-\Delta_+\log_2\Delta_+ -\Delta_-\log_2\Delta_-,
\end{eqnarray}
where $\Delta_\pm=\frac{1}{2}\left(1\pm\Gamma\right)$, $\Gamma^{2} =\left(a-d\right)^{2}+4\left(|z|+|w|\right)^{2}$, $\rho_{A}=Tr_B\rho$ is the density matrix of subsystem $A$ and $S(\rho)=-Tr(\rho\log_2\rho)$ is the von Neumann entropy. Since $S(\rho_A)=S(\rho_B)$ for the density matrix~(\ref{denmatst}), the projective measurement performed on the subsystem A or B assumes equal values. The definition of QD is emerged from the mismatch of two expressions of mutual information extended from classical to quantum systems and is basically defined as the difference between total correlations measured by mutual information and the classical correlations determined by measurement-based conditional entropy. Non-zero QD indicates the impossibility of accessing all information about one subsystem by performing a set of measurements on the other subsystem. 

Similarly, as an entanglement measure, EoF for the density matrix can be written as~\cite{PhysRevLett.80.2245}
\begin{eqnarray}\label{eof}
E=h\left[\frac{1}{2}\left(1+\sqrt{1-C^2}\right)\right],
\end{eqnarray}
where $h[x]=-x\log_2x-(1-x)\log_2(1-x)$ and $C=2\max\{0,|z|-\sqrt{ad},|w|-b\}$ is the concurrence for the density matrix~(\ref{denmatst}). EoF is one of the well motivated measure of the degree of entanglement for bipartite states. EoF and QD have the same entropic interpretation in their definitions, being equal for pure states and having strict connection by monogamic relations for mixed states. So, the two figures of merit are the natural choices for a quantitative comparison of quantum correlations. However, we should stress here that they quantify different part of quantum correlations for mixed states; there are separable mixed states with non-null QD. In fact, QD is non-zero for almost all mixed quantum states~\cite{ferraro10}. 
%%%%%%%%%%%%%%%%%%%%%%%%%%%%%%
\section{Results} \label{sec:results}
%%%%%%%%%%%%%%%%%%%%%%%%%%%%%%%
In the following, we will denote QD and EoF as $D_H$ $(D_L)$ and $E_H$ $(E_L)$ for the hot (cold) heat bath cases, respectively. We shall calculate the thermodynamical work, operational efficiency and briefly discuss the temperature effects on them as well as on quantum correlations.

\begin{figure}[ht!]
\includegraphics[width=8.0cm]{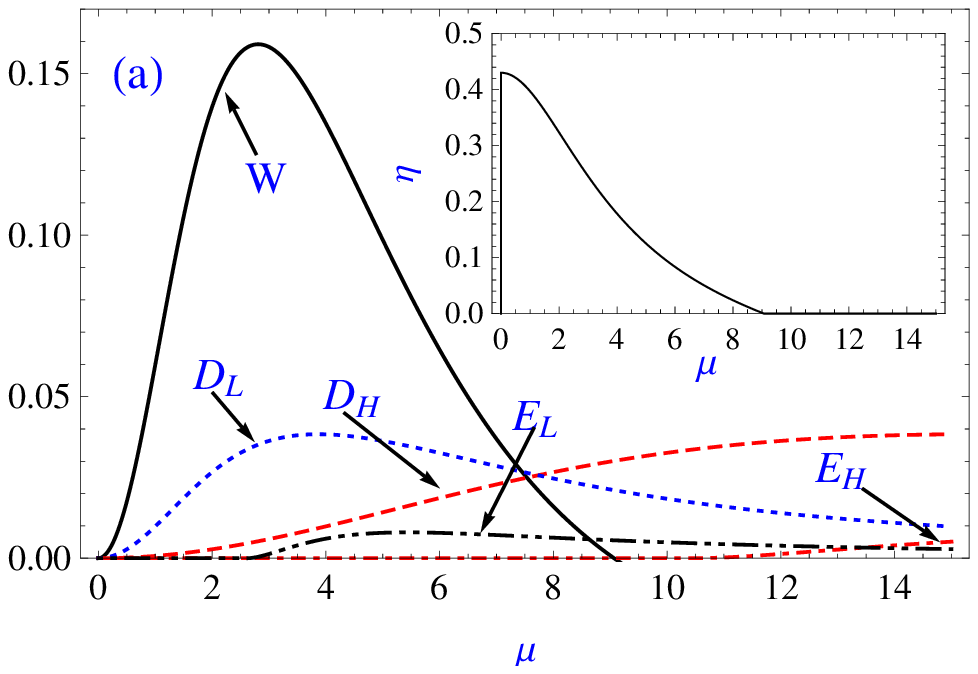}
\includegraphics[width=8.0cm]{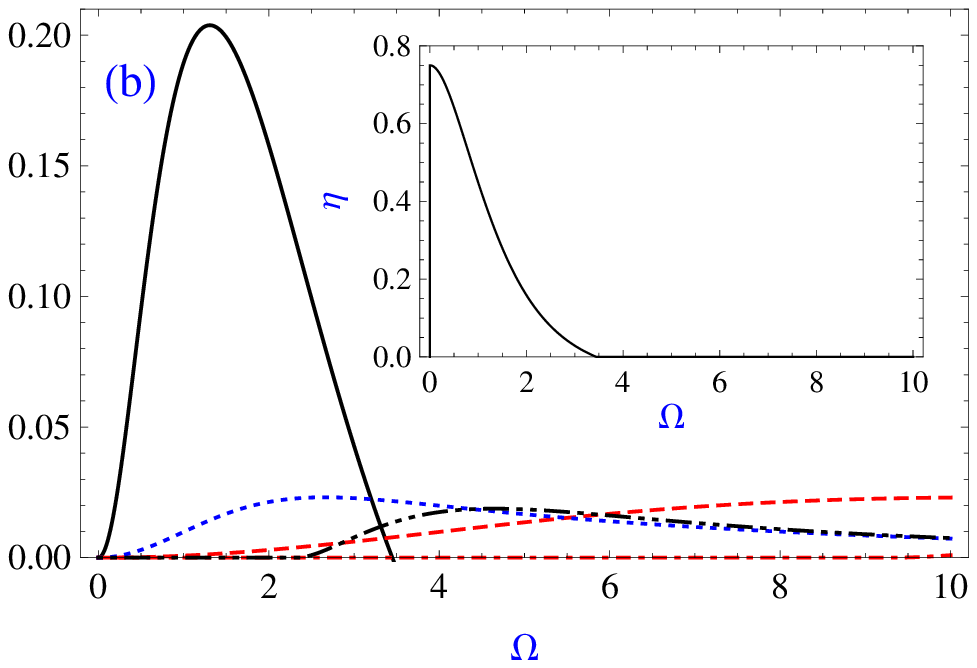}
\caption{\label{fig2}~Work (solid, line), efficiency (insets), $D_H$ (red, dashed line), $D_L$ (blue, dotted line), $E_H$ (red, dot-dashed line) and $E_L$ (black, dot-dot-dashed line) versus $\mu$ 
(a) with $\mu_H=\mu_L=\mu$, $\Omega_H=4$, $\Omega_L=1$, $k_{B}T_H=4$ and $k_{B}T_L=1$, and versus $\Omega$
(b) with $\Omega_H=\Omega_L=\Omega$, $\mu_H=4$, $\mu_L=1$, $k_{B}T_H=4$ and $k_{B}T_L=1$.}
\end{figure}

If we change the applied magnetic field ($\Omega$) in the adiabatic stages, we get different work and  efficiency out of the engine depending on the different amount of interaction ($\mu$) between the spins. We would like to see how quantum correlations build up depending on the interaction in comparison to the work output and efficiency.  In addition we ask the same question if we vary the interaction in the adiabatic stages while keeping the $\Omega$ constant. For this purpose we plot the work, operational efficiency and the quantum correlation measures for these two cases in Fig.~\ref{fig2}. In Fig.~\ref{fig2}(a) we take $\mu_H=\mu_L=\mu$, $T_H/T_L=4$ and $\Omega_H/\Omega_L=4$ and in Fig.~\ref{fig2}(b)  we take $\Omega_H=\Omega_L=\Omega$ with $T_H/T_L=4$ and $\mu_H/\mu_L=4$. We obtain from Eqs.~(\ref{spectrum}) and~(\ref{hwe}) that the efficiency and the positive work condition (PWC) for the case (a) with $\mu=0$ and for the case (b) with $\Omega=0$ have the usual formulas: (a) $\eta=1-\Omega_L/\Omega_H$ and $T_H>(\Omega_H/\Omega_L)T_L$ and (b) $\eta=1-\mu_L/\mu_H$ and $T_H>(\mu_H/\mu_L)T_L$. Therefore, the engine does not produce work for the considered parameters in Fig.~\ref{fig2} with zero squeezing (a) and magnetic field (b), since the PWC is violated. 

We see in Fig.~2(a) that the non-zero $\mu$ induces positive work. Work output first increases with $\mu$ until a critical value then drops to zero. Similarly, the non-zero $\Omega$ leads to positive work as in Fig.~\ref{fig2}(b). Changing magnetic field yields larger work and efficiency relative to changing interaction strength. The engine can operate close to the Carnot efficiency, $\eta_C=1-T_L/T_H\approx 0.75$ as shown in the inset of Fig.~\ref{fig2}(b). Efficiency decreases with $\mu$ and $\Omega$. There is optimal efficiency $\eta_o$   corresponding to maximum work which is $\eta_o\approx 0.27$ at $\mu\approx 2.7$  for case (a) and $\eta_0\approx 0.34$ at $\Omega\approx 1.25$ for case (b). The work output increases with increasing efficiency for $\eta<\eta_o$. 

At first glance, the quantum correlations, work and efficiency relations in Fig.~\ref{fig2}  have a complicated structure. On the other hand we can withdraw some general conclusions. In different models of spin systems it is reported that entanglement in the hot bath stage is detrimental for the 
work and one cannot get work if entanglement is higher in the hot bath stage than the one in cold bath stage~\cite{huang12,zhang08,zhang07,wang09}. Our engine indeed cannot produce work if it becomes entangled at the end of its contact with the hot bath; we find that $W=0$ when $E_H>0$ in Fig.~\ref{fig2}. In addition it cannot operate with high efficiency, especially close to Carnot efficiency, if it is entangled in cold bath stage, which is consistent with the results in Ref.~\cite{brunner13}. $E_L$ and $E_H$ emerge just after the critical values of $\mu$ and $\Omega$. 

On the other hand, quantum discord always exists and our engine can actually work with high efficiency, 
even close to Carnot efficiency, with quantum correlations beyond entanglement. Fig.~\ref{fig2} shows that 
work can be extracted when $D_H>0$ for both cases and even when $D_H>D_L$ in case (a). $W$ and $D_L$ have nearly monotonic relation for the case (a). Entanglement in the cold bath stage is not detrimental for the operation of the engine, though it is not strong as much as quantum discord. We can conclude that the quantum correlations beyond entanglement both in cold and heat bath cases can play a constructive role on the performances of the heat engine. Following analysis will explore this conclusion further  and at the end we will focus more on the positive effects of quantum discord in the cold bath stage on the efficiency and work extraction.

\begin{figure}[ht!]
\includegraphics[width=8.0cm]{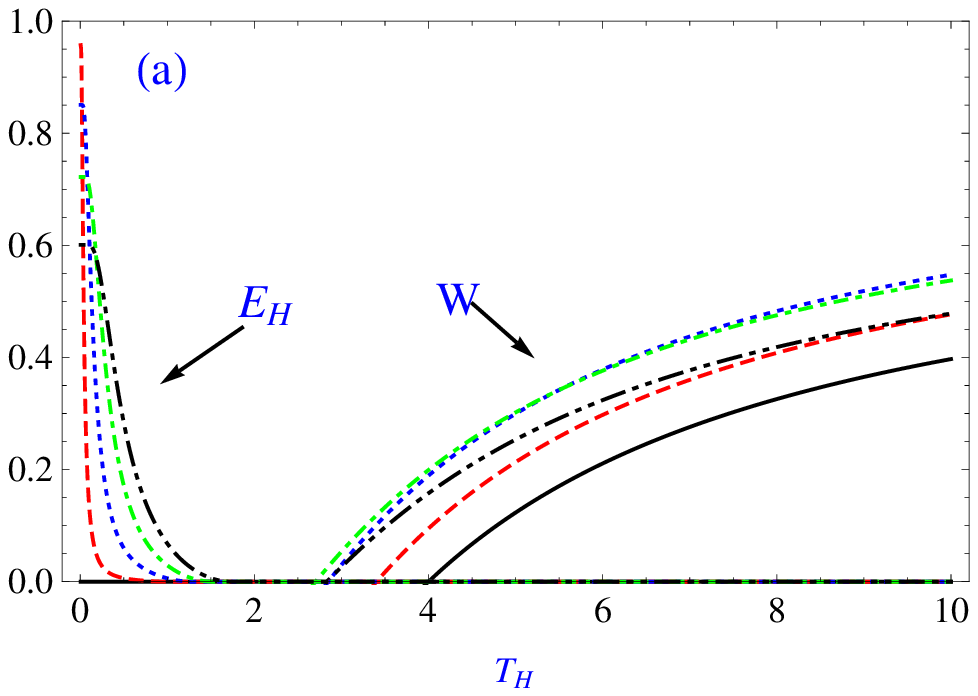}
\includegraphics[width=8.0cm]{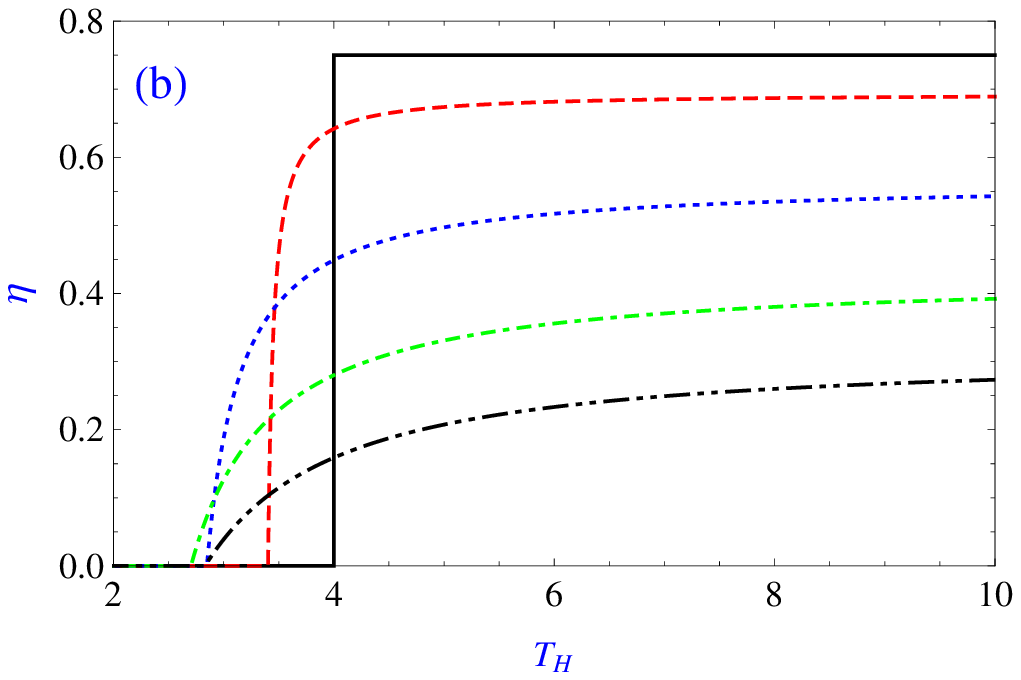}
\caption{\label{fig3} Dependence of entanglement of formation (EoF) $E_H$~((a), right bunch), positive work $W$~((a), left bunch) and efficiency $\eta$~(b) on the temperature of the hot reservoir $T_H$ (in units of $k_{B}$) for $k_{B}T_L=1$, $\mu_H=4$, $\mu_L=1$, $\Omega=0$~(black, solid line), $\Omega=0.5$~(red, dashed line), $\Omega=1$~(blue, dotted line), $\Omega=1.5$~(green, dot-dashed line) and $\Omega=2$~(black, dot-dot-dashed line). Note that for $\Omega=0$, EoF and quantum discord are both zero in hot and cold bath stages.}
\end{figure}

We analyze the effect of the hot bath temperature and external field on the positive work, efficiency and the quantum correlations in the case of changing interaction strength during the adiabatic stages.  We have also performed a similar analysis for the case where external field is changed at the adiabatic stages and found the similar qualitative results. Therefore, those results are not displayed in the text. Fig.~\ref{fig3} shows that both the work and efficiency increase with $T_H$.  We see from Fig.~3(a) that the monotonic increase in work with $T_H$ is in an expected form. Higher $T_H$ for a given $T_L$ leads to more heat absorption, and more work is produced. External field increases the work output relative to $\Omega=0$ case as in Fig.~\ref{fig3}(a), while it lowers the efficiency and 
makes it slightly temperature dependent as in Fig.~\ref{fig3}(b). $W$ increases up to a critical point then decreases with $\Omega$. Fig.~\ref{fig2} is a special case of this general behavior observed in Fig.~\ref{fig3}(a) where there is a critical $\Omega$ for which $W$ is maximum for given $T_H$ and $T_L$. The engine operates with a fixed $\eta=1-(\mu_L/\mu_H)$ at $\Omega=0$ and produces work just after the PWC, $T_H>(\mu_H/\mu_L)T_L$. The external field modifies the PWC and makes the engine operate at a lower $T_H$ compared to the $\Omega=0$ case.

The quantumness is expected to degrade with the temperature. Fig.~\ref{fig3}(a) confirms with this expectation. $E_H$ decreases with $T_H$ and disappears at $T_H\approx 2$. $E_L$ is always zero in $0\leq\Omega\leq 2$. We conclude that the working substance is fully separable at the end of two heat bath stages. On the other hand, quantum correlations beyond entanglement can survive at higher temperatures. 
We calculated $D_H$ and $D_L$ and find that they can coexist with the positive work. In the parameter regime of Fig.~\ref{fig3}(a), $D_H$ behaves qualitatively the same as $E_H$ while $D_L$ is constant. Both are small $D_H<0.01$, $D_L<0.02$ but nonzero in the positive work regime. Fig.~\ref{fig2} is an example of his situation at a particular set of $T_H$ and $T_L$. 

\begin{figure}[ht!]
\includegraphics[width=8.0cm]{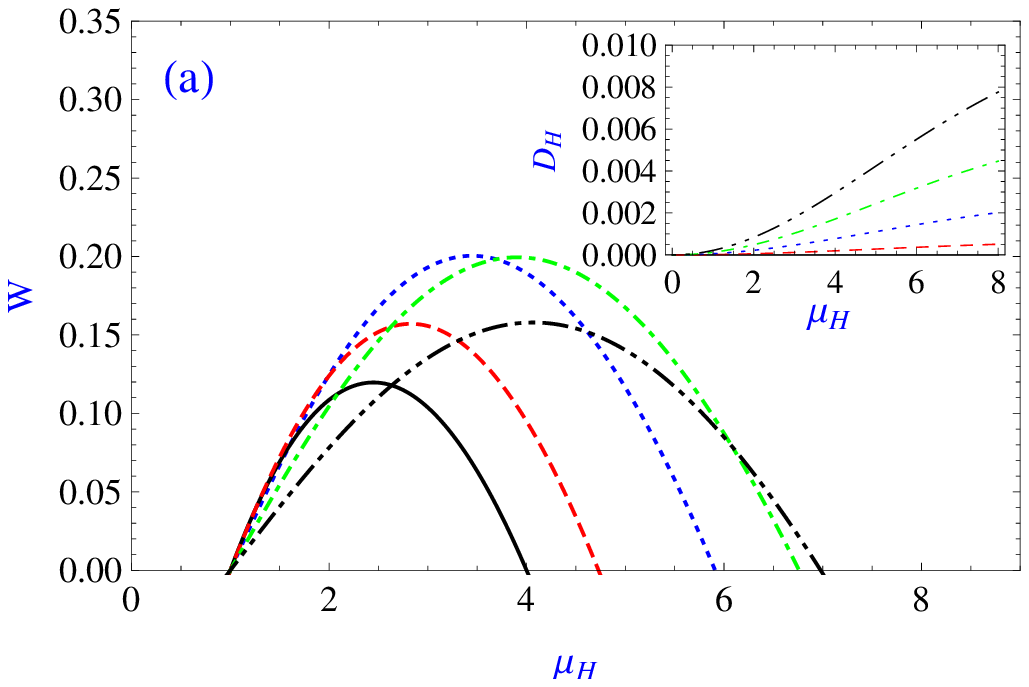}
\includegraphics[width=8.0cm]{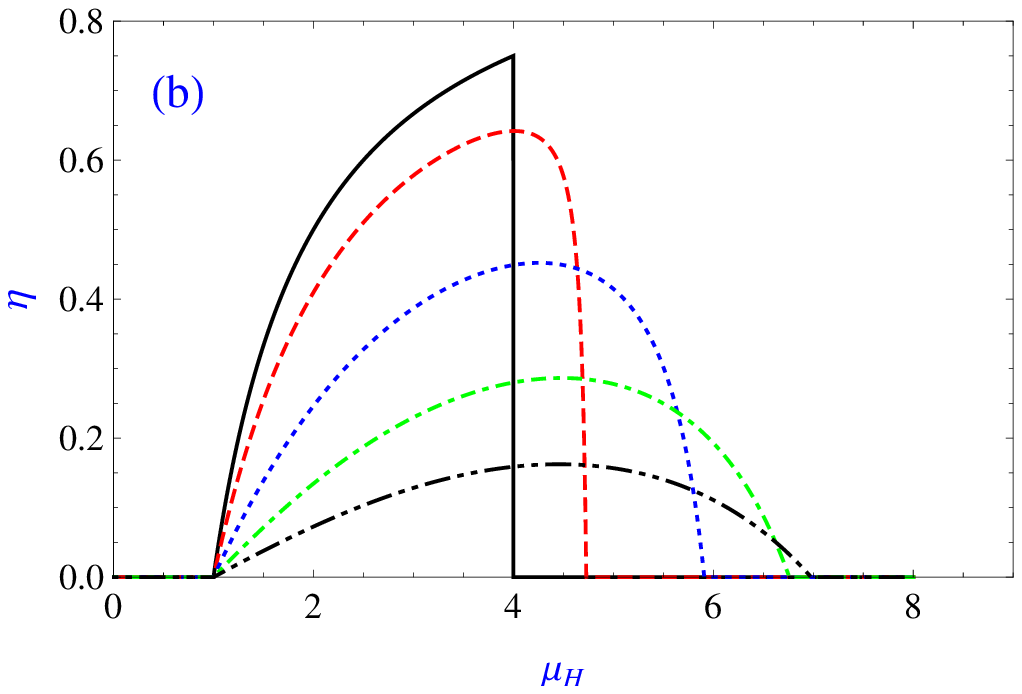}
\caption{\label{fig4} Dependence of positive work $W$~(a) and efficiency $\eta$~(b) on $\mu_H$ for $k_{B}T_H=4$, $k_{B}T_L=1$, $\mu_L=1$, $\Omega=0$~(black, solid line), $\Omega=0.5$~(red, dashed line), $\Omega=1$~(blue, dotted line), $\Omega=1.5$~(green, dot-dashed line) and $\Omega=2$~(black, dot-dot-dashed line). The inset in (a) shows $D_H$ versus $\mu_H$ for the same parameters. Note that for $\Omega=0$, entanglement of formation  and quantum discord are zero for both hot and cold bath stages.}
\end{figure}

Next, we analyze the effect of $\mu_H$ and $\Omega$ in the work extraction and the efficiency of the engine depicted in Fig.~\ref{fig4}(a) and Fig.~\ref{fig4}(b), respectively. If there is no external field, the strict conditions, $\mu_H>\mu_L$ and $T_H>(\mu_H/\mu_L)T_L$, determine the operation regime of the engine as $\mu_L<\mu_H<4\mu_L$  in Fig.~\ref{fig4}(a). The interval gets larger with $\Omega$. 
The efficiency increases until the Carnot point with $\mu_H$. In the region, $1<\mu_H<2.5$, the work output increases with the increasing efficiency. The non-zero $\Omega$ can lead to the increase the work output at the cost of reduced efficiency. 

For the temperature ranges in Fig.~\ref{fig4}, the working medium has no entanglement ($E_H=E_L=0$) at the end of the stages $1$ and $3$ of the engine cycle, while the quantum correlations beyond entanglement exist. $D_H$ increase with the interactions as shown in the inset of Fig.~\ref{fig4}(a) while $D_L$ is independent of $\mu_H$ and hence a non-zero constant depending on $\Omega$. There are critical amount of quantum correlations indicated by $D_H$ at the maximum $W$ and $\eta$. Work output is larger but at a reduced efficiency relative to uncorrelated engine (QD and EoF are both zero.) with $\Omega=0$. The quantum correlated Otto engine, with dominant correlations measured by QD, produces work in the
parameter regimes ($\mu_H>4$) where the uncorrelated engine ($\Omega=0$) cannot.
 Similar analysis for the  case where $\Omega$ is changed at the adiabatic stages produce similar results, and hence we have not presented them in the text.
\begin{figure}[ht!]
\includegraphics[width=8.0cm]{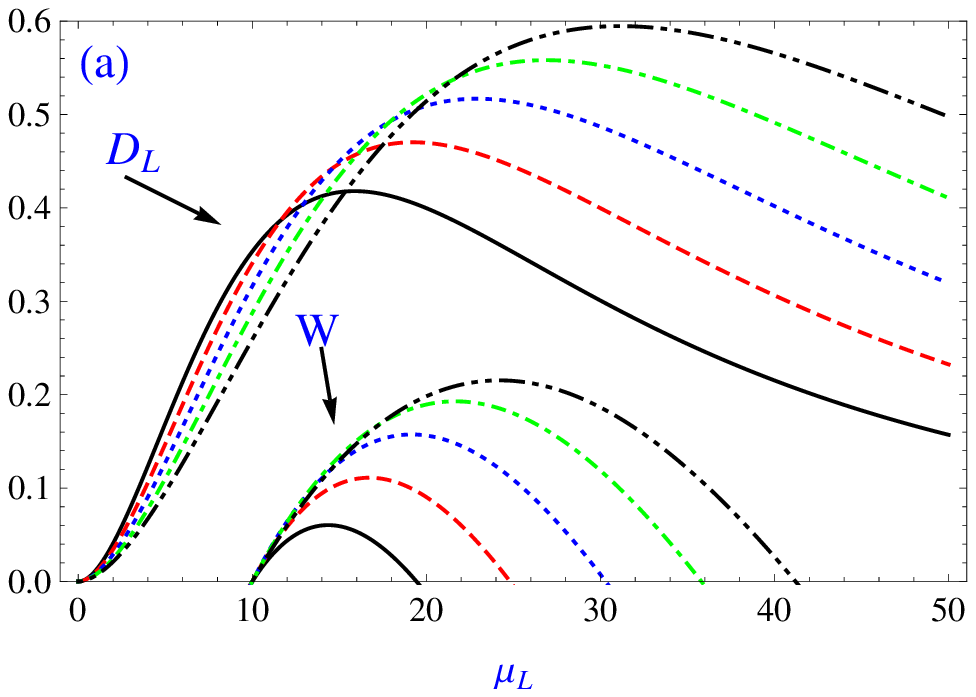}
\includegraphics[width=8.0cm]{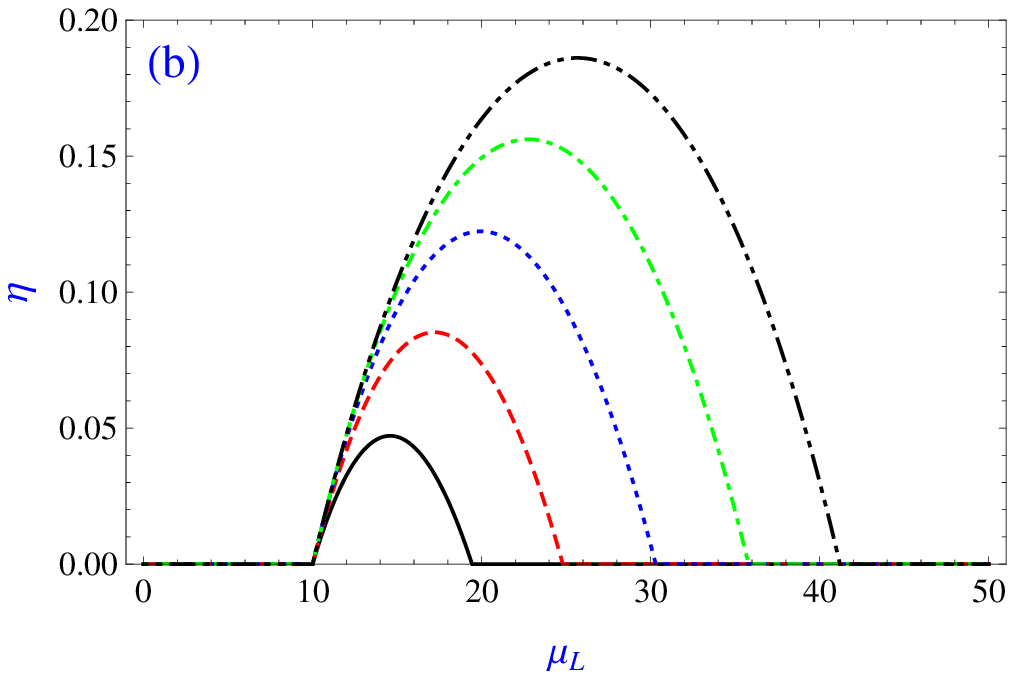}
\caption{\label{fig5} Dependence of positive work $W$~((a), bottom bunch), $D_L$~((a), top bunch) and efficiency $\eta$~(b) on $\mu_L$ for $k_{B}T_H=4$, $k_{B}T_L=1$, $\mu_H=10$, $\Omega=7$~(black, solid line), $\Omega=8$~(red, dashed line), $\Omega=9$~(blue, dotted line), $\Omega=10$~(green, dot-dashed line) and $\Omega=11$~(black, dot-dot-dashed line). Note that $E_L$ is qualitatively same as $D_L$, so it is not plotted here. Also note that $D_H$ and $E_H$ increase as $\Omega$ increases.}
\end{figure}

The energy gaps of the working substance play a subtle role in the performances of the quantum heat engines and determine the conditions in which work can be extracted efficiently~\cite{quan07,thomas11}. For instance, if the system is non-interacting $\mu=0$, we have $E_{1,4}=\mp\Omega$ while $E_2=E_3=0$. The energy gaps get larger if $\Omega$ is increased from $\Omega_H$
to $\Omega_L$ in the second adiabatic stage of the Otto cycle. This would correspond to adiabatic contraction process, instead of expansion, and hence the engine could not produce work. This complies with the requirement of $\Omega_L<\Omega_H$ by the efficiency expression $\eta=1-(\Omega_L/\Omega_H)$. Similarly, if there is no external field $\Omega=0$, then the level structure becomes
$E_{1,3,4}=\mu$ and $E_2=0$. Increasing the interaction strength from $\mu_H$ to $\mu_L$ in the second stage again make the
adiabatic process contraction rather than an expansion. The efficiency $\eta=1-(\mu_L/\mu_H)$ become negative and the engine cannot
produce work. Presence of external field or interactions can change this picture during the variation of the control parameters in the adiabatic stages.

A natural question which could arise is the possibility of work extraction when $\Omega_H<\Omega_L$ in the case of interacting working substance with non-zero $\mu_H=\mu_L=\mu$, and when $\mu_H<\mu_L$ in the case of non-zero external field $\Omega_H=\Omega_L=\Omega$. The former case is not possible, since non-zero $\mu$ introduces just a shift to the energy levels, and the structure of the energy gaps remains the same. For the latter case, we can find narrowing gaps which are $E_2-E_1$ and $E_4-E_3$,
as can be seen in Fig.~\ref{fig1}(b), as $\mu_H$ is increased to $\mu_L$ in the stage $2$ of the cycle. For the small temperature limit or for large system parameters, the levels $E_1$ and $E_2$ are well separated from the others and contribution of their gap to work extraction is dominant. Accordingly, it would be possible to harvest positive work from the engine for $\mu_H<\mu_L$. Moreover, this effect arises in the deep quantum regime of the strongly interacting working medium so that the quantum correlations are expected to play a decisive role. Indeed the maximum of the QD, positive work and efficiency appear almost at the same $\mu_L$ in Fig.~\ref{fig5}; beyond which they all decrease. Presence of non-zero $\Omega$ increase the QD, work and efficiency. The machine produces work efficiently for the case $\mu_L>\mu_H$ ($\mu_H=10$ in Fig.~\ref{fig5}.) where this situation cannot be achieved in the absence of magnetic field. Similar interplay of QD, work and efficiency is  found for the EoF, $E_L$, as well.
%%%%%%%%%%%%%%%%%%%%%%%%%%%%%%%%%%%%%%%%%%%%%%%%%%%%%%%%
\section{Conclusions}\label{sec:conc}
%%%%%%%%%%%%%%%%%%%%%%%%%%%%%%%%%%%%%%%%%%%%%%%%%%%%%%%%
In summary, we proposed a four-level, correlated quantum Otto engine based on a one-axis twisting spin squeezing model with an external magnetic field. We discussed the work extraction from the engine and its efficiency as well as the role of quantum correlations, characterized by the quantum discord (QD) and the entanglement of formation, in the thermodynamical processes. 

In our interacting spin Otto engine, a quantum working substance (quantum fuel) is generated by a particular nonlinear spin-spin interaction known as axis-twisting spin squeezing interaction. This interaction yields quantum states of the bipartite spin system having pairwise entanglement as well as quantum correlations beyond entanglement, which can be characterized by a measure called quantum discord. 

The quantum fuel is cooled to low temperatures and compressed by a quantum adiabatic process in the preparation stages of the Otto cycle. These are described as the third and the fourth stages in our case. Compression means larger energy gaps for the quantum fuel. At low temperatures with strong interactions the quantum fuel is prepared with a quantum discord $D_L$.
The cycle continues with the stages one and two where the fuel is burnt and expanded by another quantum adiabatic process to yield positive work. The expansion stage decreases the energy gaps. The system looses its “quantumness” and increases its classicality. 
When the quantum fuel is burnt, its quantum discord is reduced to $D_H$. This could be interpreted as using the quantum information gradient as a resource, similar to the thermal gradient, to be further harvested by the engine 
to enhance positive work. 

We find that whether QD or entanglement can be used as a resource or if they are just byproducts depend on the parameter regimes. We can generally conclude for our model that entanglement of the working substance in the hot bath, $E_H$ forbids the work extraction if the external field or the interactions are stronger in hot bath stage relative to cold bath stage. In this case $D_L$ is constructive over a wide range of interaction strengths while $D_H$ can be either constructive or destructive depending on the system parameters. Their difference, $D_L-D_H$ on the other hand is constructive as long as $D_L>D_H$. When the interaction is weaker in the hot bath stage than the one in cold bath stage, both the entanglement and QD, irrespective of whether they are in the hot or cold bath stages, lead to enhanced work and efficiency. 

Our results suggest that choosing a suitable interaction model and optimizing system parameters are crucial to exploit quantum correlations as a resource. Our model of nonlinear two-spin quantum Otto engine is a special example, which can inspire further general or specific studies of designing optimal quantum machines.

%%%%%%%%%%%%%%%%%%%%%%%%%%%
\acknowledgements
The authors warmly thank N.~Allen for inspiration. F.~A.~acknowledges the support and hospitality of the Off{\.i}ce of Vice President for Research and Development (VPRD) and Department of Physics of the Ko\c{c} University. \"O.~E.~M.~ acknowledges the support by Lockheed Martin Corporation Grant. A.~\"U.~C.~H. acknowledges the COST Action MP1209.
%%%%%%%%%%%%%%%%%%%%%%%%%%%%%%%%%%%%%%%%%%%%%%%%%%%%%%%%%%%%%%%%%%%%%%%%%%%%%%
%\bibliographystyle{apsrev}
%\bibliography{S_Sq_QHE}{}
%%%%%%%%%%%%%%%%%%%%%%%%%%%%%%%%%%%%%%%%%%%%%%%%%%%%%%%%%%%%%%%%%%%%%%%%%%%%%%

\end{document}